\documentclass[aps,prl,epsfig,floats,twocolumn,superscriptaddress,amssymb,amsmath,floatfix,showpacs]{revtex4}
\usepackage{graphicx}
\usepackage{bm}  % bold math
\usepackage{amssymb}
\usepackage{color}
\usepackage{amscd}

\begin{document}

%\title{Molecular Hunter: a Theoretical Approach to Bacterial Chemotaxis}
%\title{Stochastic Molecular Hunter}
%\title{Bacterial Chemotaxis: a Basic Mechanism for a Molecular Hunter}
\title{Hydrodynamics of a Micro-Hunter: Chemotactic Scenario}

\author{Ali Najafi}
\email{najafi@znu.ac.ir} \affiliation{Department of Physics,
Zanjan University, Zanjan 313, Iran}

\date{\today}

\begin{abstract}
Inspired by bacterial chemotaxis we propose a hydrodynamic molecular scale hunter that can swim and find its target. The system is essentially a stochastic low Reynolds swimmer with ability to move in two dimensional space and sense the local value of the chemical concentration emitted by a target.
We show that by adjusting the geometrical and dynamical variables of the swimmer we can always achieve a swimmer that can 
navigate and search for the region with higher concentration of a chemical emitted from a source.
The system discussed here can also be considered as a theoretical framework for describing the 
bacterial chemotaxis.
\end{abstract}
%07.10.Cm Micromechanical devices
%81.07.Nb 	Molecular nanostructures
%82.39.-k 	Chemical kinetics in biological systems
%87.19.ru 	Locomotion
%87.17.Jj 	Cell locomotion, chemotaxis 
%47.15.G- 	Low-Reynolds-number (creeping) flows 
%45.40.Ln 	Robotics

\date{\today}

\maketitle
Propulsion mechanisms for microorganisms and artificial swimmers are subject to the 
exceptional constraints of the motion in low Reynolds number hydrodynamics
\cite{PoLa-rev,Purcell}. Purcell's Scallop theorem beautifully demonstrates how 
a set of non-reciprocal body deformations is necessary to achieve a  net translational or rotational 
movements in simple systems \cite{becker,3SS,josi,drey,feld,peko,yeomans1,Karsten}. 
Experimental verification of the swimming motion in systems with only a small number of internal degrees 
of freedom have attracted interest in developping new artificial swimmers \cite{exp-swim,exp-swim2}.
In biological systems, sperm cell as micron scale hunter,  uses beating flagella to swim toward its targets \cite{Kruse,bray,spermchem,berg2}. 
These targets include food that is necessary for surviving and the egg cell that is essential in 
fertilization process. A concentration gradient of chemical emitted by the source is established in this 
chemotaxis phenomena \cite{BergEC}.
The physical mechanism of chemotaxis in flagellated cells like sperm, with circling trajectories 
are usually described in the following way \cite{frankchem}: the underlying chemichal network of 
an active  stimulus-response system provides a concentration mediated stimulus which 
periodically regulates the internal motion and  modulate the curvature of the swimming path. 
The input in this signaling system is the local value of a chemical \cite{stimulres1,stimulres2}. 
This scenario gives rise to a drift in the circular trajectories of  chiral flagellated swimmers \cite{chiralswimmer,frankchem}.

\begin{figure}
  \centerline{\includegraphics[width=.60\columnwidth]{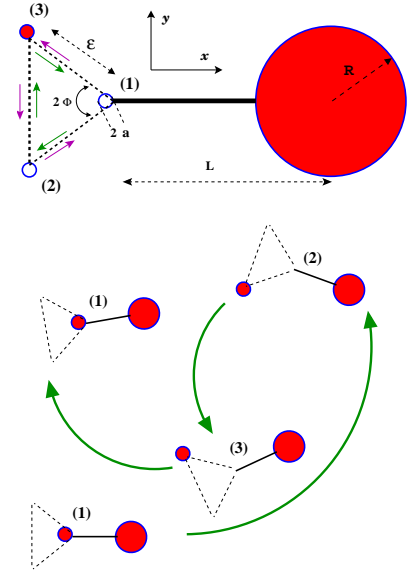}}% Images in 100% size
  \caption{(color online). Two large and small spheres are connected through an arm to construct a swimmer that can move 
in (2-D) space. This model resembles the 
geometry of a bacterium that has a single tail. In a stochastic description of the motion 
we assume that the small sphere can be in one of the three distinct states as shown in the picture.  
The overall motion of the system in a complete cycle of the stochastic jumps are shown.}
\label{fig1}
\end{figure}
In this letter, inspired by chemotaxis we propose a molecular scale swimmer that can navigate in 
 (2-D) space, along the 
gradient of a stimulating chemical. We assume that the chemical can activate a simple 
relaxational process in the swimmer and correspondingly initiates signals to change the internal 
motion leading the system to find the right track.  
The effects of fluctuations have been taken into account by considering a stochastic description 
of the internal motion of the swimmer. We investigate the conditions that  
the swimmer can reach the region with higher concentration of the chemical. The model presented here can also be regarded as a theoretical  model to  account the hydrodynamic details of chemotaxis for a chiral 
swimmer with a single flagellum.

Consider  a minimal  hydrodynamic propeller composed of two large and small spheres with radii $R$ and 
 $a$ where $a< R$. These spheres are connected through a negligible diameter arm that is not 
interacting with the ambient fluid. Different internal configurations of the system can be achieved 
by changing the length and the shape of the arm. As a specific choice we consider the case where the arm 
always stay in a two dimensional plane, as shown schematically in Fig. \ref{fig1}. 
For further simplification of the model we decrease the number of internal configurations to three 
different states denoted by states $(1)$, $(2)$ and $(3)$.  State $(1)$ is a reference state where two 
spheres are separated by a distance $L$. States $(2)$ and $(3)$ are characterized by a distance $\epsilon$ 
and an angle $\Phi$ with respect to the reference state $(1)$, as depicted in Fig. \ref{fig1}. 
The small sphere can jump between the denoted states along the shown paths with a constant 
velocity  $v_0$ measured in the reference frame of the large sphere. A complete cycle of the motion 
fulfills the Purcell's Scallop theorem  and eventually leads the system to a new state with 
a net translational and rotational displacements \cite{Purcell}. The motion is restricted to a (2-D) plane that is characterized by the plane of the arm.

To analyze the hydrodynamics of the system, we use the fact that the linear Stokes equation governs the 
hydrodynamics of the fluid flow at zero Reynolds number. 
To simplify the description, we furthere assume that the radius of small sphere is much smaller than the other length scales of 
the system, namely $R$ and $L$ ($a\ll R$, $a\ll L$). In this case the 
motion of the small sphere can be regarded  as a singular body force located at the position of this 
sphere. Taking the advantage of this simplification and using the linearity of the Stokes equation,
we can describe the dynamics of the system for a general internal motion. 
In the reference frame, that is co-moving and rotating with large sphere, we denote the position vector 
of the small sphere by ${\bf r}_0(t)$. 
Denoting the linear and angular velocities of the large sphere in a laboratory frame 
by ${\bf V}$ and ${\bf \Omega}$, we can express the velocity field of the fluid at a general point 
${\bf r}$ in the co-moving frame as: 
\begin{equation}
{\bf u}({\bf r})=-{\bf V}-{\bf \Omega}\times {\bf r}+ M.{\bf V}+{\bf \Omega}\times {\bf m}+ 
G({\bf r},{\bf r}_0).{\bf f},
\end{equation}
where ${\bf f}$ denotes the strength of the point force located at the position of the small sphere.
Here the tensor $M$ and vector ${\bf m}$ give the flow field due to the translational and rotational motion of a moving sphere and are given explicitely as: $M=\frac{3}{4}\frac{R}{r}[{\bf I}+\frac{{\bf r}{\bf r}}{r^{2}}]+\frac{1}{4}\frac{R^{3}}{r^{3}}[{\bf I}-3\frac{{\bf r}{\bf r}}{r^{2}}]$ and ${\bf m}=\frac{R^{3}}{r^{3}} {\bf r}$. The Green's function of the Stokes equation for an infinite flow 
bounded internally by a solid sphere with radius $R$ is denoted by $G({\bf r},{\bf r}_0)$. The explicit form 
of this Green's function has been calculated by C.W. Oseen \cite{Oseen}.

Force and torque balances for a self propeller system require that the total force and torque acting on the 
fluid should vanish. The point force located near the sphere has an image force with strength 
${\bf f}_{i}=(c_{1}{\hat {\bf r}}_0{\hat {\bf r}}_0+c_{2}{\bf I})\cdot{\bf f}$ with $c_1=-\frac{3}{4}[\frac{R}{r_{0}}-\frac{R^{3}}{r_{0}^{3}}]$ and $
c_2=-\frac{3}{4}\frac{R}{r_{0}}-\frac{1}{4}\frac{R^{3}}{r_{0}^{3}}
$ and ${\bf I}$ is the unit matrix. This image force is located inside the sphere in position given by: ${\bf r}_{0}^{*}=\frac{R^2}{r_0^2}{\bf r}_0$.
A discussion by Higdon shows that in evaluating the force acting on the fluid one should carefully 
account the image system \cite{Hig}. In this case the force and torque balances read:
\begin{equation}
{\bf f}+{\bf f}_{i}+6\pi\eta R {\bf V}=0,~~
{\bf r}_{0} \times {\bf f}+{\bf r}_{0}^{*}\times {\bf f}_{i}+8\pi\eta R^{3}{\bf \Omega}=0.\nonumber
\end{equation}
To finish with the dynamical equations we should include the prescribed form of the internal motion by the boundary condition: ${\bf u}({\bf r}_0)=\dot{{\bf r}}_0$.  Having in hand the force balance equations and 
this boundary condition, we can eliminate the point force strength and arrive at equations for linear and 
angular velocities of the system and obtain velocity of large sphere as: 
${\bf V}={\bf A}\cdot\dot {{\bf r}}_0$ and 
${\bf \Omega}={\bf B}\cdot\dot {{\bf r}}_0$, where ${\bf A}$ and $\bf B$ are two matrices with elements that strongly depend on the specific 
form of the internal motion given by function ${\bf r}_0(t)$ \cite{2S}. 
Here instead of giving the explicit expression of the swimming velocity for a general motion, 
we concentrate on the simple motion of our (2-D) model that has been introduced and discussed already.  

We denote the dynamical variables of the system by ${\bf x}$ and $\theta$, where ${\bf x}$ 
stands for the  position vector 
of the large sphere and $\theta$ measure the angle that the  swimmer's director makes with the $x-$axis. 
Swimmer's director is defined as a unit vector pointing from the position of the reference state $(1)$ to the 
center of the large sphere. Note that $\dot{{\bf x}}={\bf V},~\dot{\theta}={\bf \Omega}$. Now the differential changes of the swimmer's variables in a general jump from state $(i)$ to state $(j)$ can be written as:
\begin{equation}
\Delta{\bf x}_{ij}={\bf R}^{-1}(\theta)\cdot{\bf d}_{ij},~~~
\Delta\theta_{ij}=\alpha_{ij},
\end{equation}
where ${\bf R}(\theta)$ represents the matrix for a rotation around $z-$axis by the 
instantanous value of the angle $\theta$. 
Let us consider the case where the internal deformations are small, compared to the average 
length of the swimmers, such that $\epsilon\ll L$. This allows us to set up a perturbative expansion 
of the results. Up to the leading order in $\epsilon$ and $a$, the differential rotations read:
\begin{equation}
\alpha_{12}=-\frac{3}{4}\left(\frac{\epsilon}{R}\right)\left(\frac{a}{R}\right)\left(1+\frac{L}{R}\right)\sin\Phi,
\end{equation}
$\alpha_{23}=-4\alpha_{12}$ and $\alpha_{31}=\alpha_{12}$.
%To the leading order in $\epsilon$, the displacement vectors read: ${\bf d}_{12}=\begin{pmatrix} \delta_1 \\ %\delta_2 \end{pmatrix}$, 
%${\bf d}_{23}=\begin{pmatrix} 0 \\ \delta_3 \end{pmatrix}$ and 
%${\bf d}_{31}=\begin{pmatrix} -\delta_1 \\ \delta_2 \end{pmatrix}$
% where, 
%\begin{eqnarray}
%&&\delta_1=\epsilon\left(\frac{a}{R}\right)\left(1-\frac{3}{2}\frac{R}{L}\right)\cos\Phi,\nonumber\\
%&&\delta_2=
%\epsilon\left(\frac{a}{R}\right)\left(1-\frac{3}{4}\frac{R}{L}\right)\sin\Phi,
%\end{eqnarray}
To the leading order in $\epsilon$ and $a$, the displacement vectors read: 
%\begin{equation}
$$
{\bf d}_{12}=\begin{pmatrix} \delta_1 \\ \delta_2 \end{pmatrix},~~ 
{\bf d}_{23}=\begin{pmatrix} 0 \\ \delta_3 \end{pmatrix},~~
{\bf d}_{31}=\begin{pmatrix} -\delta_1 \\ \delta_2 \end{pmatrix},
$$
where
%\end{equation}
%where, $\delta_1=\epsilon\left(\frac{a}{R}\right)\left(1-\frac{3}{2}\frac{R}{L}\right)\cos\Phi$, $\delta_2=
%\epsilon\left(\frac{a}{R}\right)\left(1-\frac{3}{4}\frac{R}{L}\right)\sin\Phi$ 
%\begin{eqnarray}
%&&\delta_1=\epsilon\left(\frac{a}{R}\right)\left(1-\frac{3}{2}\frac{R}{L}\right)\cos\Phi,\nonumber\\
%&&\delta_2=
%\epsilon\left(\frac{a}{R}\right)\left(1-\frac{3}{4}\frac{R}{L}\right)\sin\Phi,
%\end{eqnarray}
\begin{equation}
\delta_1=\epsilon\frac{a}{R}(1-\frac{3}{2}\frac{R}{L})\cos\Phi,~~
\delta_2=
\epsilon\frac{a}{R}(1-\frac{3}{4}\frac{R}{L})\sin\Phi,\nonumber
\end{equation}
and $\delta_3=-4\delta_2$ and the results are given for $R\ll L$.
Scallop theorem allows us to simply express the changes for reverse jumps in terms of the forward jumps 
such that: $\Delta{\bf x}_{ji}=-\Delta{\bf x}_{ij}$ and $\Delta\theta_{ij}=-\Delta\theta_{ji}$. 
We have assumed that all jumps happen with a constant velocity $v_0$. In this case the 
time for the jump $1\rightarrow 2$ ($3\rightarrow 1$) is equal to $\tau_1=\epsilon/v_0$ ($\tau_3=\epsilon/v_0$) and the time for the jump $2\rightarrow 3$ is equal to 
$\tau_2=2\sin\Phi\epsilon/v_0$. The trajectory of the motion for the swimmer moving in a cyclic way 
(1$\rightarrow$2$\rightarrow$3$\rightarrow$1) is a circular path with curvature given by: 
$\kappa_{0}^{-1}=\left(2\delta_2+\delta_3\right)$. In this motion the swimmer's director rotates with 
an angular frequency given by: $\Omega_0=\left(2\alpha_{12}+\alpha_{23}\right)/\left(2\tau_1+\tau_2\right)$.

To consider the effects of fluctuations which are present in any physical system, we  develop a 
stochastic description of the system, where the jumps between different states can happen with different 
rates. To construct the the stochastic model we denote the probability of the system to be in the 
state  $(i)$ by $P_i$. The transition rate for stochastic jump from state $(i)$ to state $(j)$ 
is denoted by $\omega_{ij}$. The rates for internal conformational changes 
in general depends on the temperature of the fluid and the detail internal activity of the system. 
Dynamics of this stochastic system is governed by the Fokker-Planck equations as:
$\dot{P}_{1}=\omega_{31}P_{3}+\omega_{21}P_{2}-(\omega_{12}+\omega_{13})P_{1},~
\dot{P}_{2}=\omega_{32}P_{3}+\omega_{12}P_{1}-(\omega_{21}+\omega_{23})P_{2}$, where 
dot symbol denotes the time derivative and also note that the probability conservation implies that $P_{1}+P_{2}+P_{3}=1$. 
The differential change of the displacement  per unit time for this stochastic system 
can be written as:
\begin{eqnarray}
&&\frac{d{\bf x}}{dt}=P_{1}\left(\omega_{12}\Delta{\bf x}_{12}+\omega_{13}\Delta{\bf x}_{13}\right)+
\nonumber\\&&
P_{2}\left(\omega_{23}\Delta{\bf x}_{23}+\omega_{21}\Delta{\bf x}_{21}\right)
+
P_{3}\left(\omega_{31}\Delta{\bf x}_{31}+\omega_{32}\Delta{\bf x}_{32}\right),\nonumber
\end{eqnarray}
and a similar equation for the rate of change of $\theta$ while we replace all $\Delta{\bf x}_{ij}$ with $\alpha_{ij}$.
For a system that is in the thermodynamic equilibrium, the transition rates for all jumps are symmetric ($\omega_{ij}=\omega_{ji}$) and the time averaged velocity of this system is zero. If for any reason 
the detailed ballance violates in the internal conformational changes, 
the system will behave like a circle swimmer. As an example let us assume that the rates for all 
the  clockwise 
jumps are equal, such that $\omega_{12}=\omega_{23}=\omega_{31}=\omega_{0}$ and all the other 
counterclockwise jumps set to $\omega_{0}+\delta\omega$. In this case and for small $\delta\omega$ the trajectory is a circle with radius of curvature given by $\kappa^{-1}=(\delta\omega/\omega_0)\kappa_{0}^{-1}$, 
where we have already defined $\kappa_0$.
\begin{figure}
  \centerline{\includegraphics[width=.90\columnwidth]{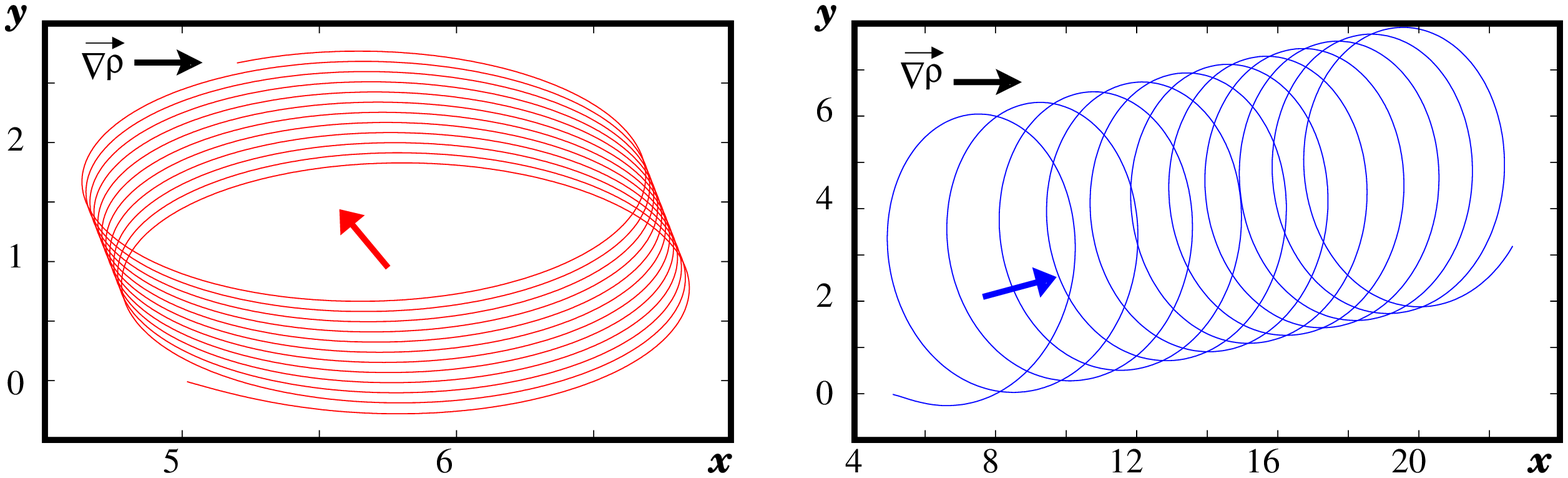}}% Images in 100% size
  \caption{(color online). Two different trajectories 
for the swimmer moving in a linear gradient of concentration 
given by $\rho({\bf x})=a_1+a_2x$ are shown. Parameters in both graphs are given by: 
$R=1,~\omega_0=1,~a=0.1,~L=3.1,~\epsilon=0.4,~a_1=2,~a_2=0.5,~\sigma=1,~\mu=10$, 
for left graph: $\Phi=0.2\pi,~\omega_{13}=1.5$ while for right graph: $\Phi=0.1\pi,~\omega_{13}=1.5$. 
For initial conditions we have chosen: $x(0)=5,~y(0)=0,~\theta(0)=0$.
Two different behavior of the trajectories can be distinguished: motion toward higher or lower concentrations.
}\label{fig2}
\end{figure}

\begin{figure}
  \centerline{\includegraphics[width=.90\columnwidth]{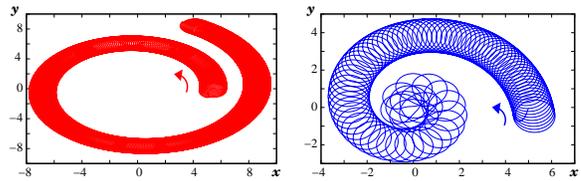}}% Images in 100% size
  \caption{(color online). Two different trajectories for 
the swimmer moving in a central gradient of concentration 
given by $\rho({\bf x})=a_3/|{\bf x}|$ are shown. Parameters for left graph 
are: $\Phi=0.2\pi,~\omega_{13}=0.8$ and for right graph: $\Phi=2.6\pi,~\omega_{13}=0.5$, other parameters are the same parameters used in Fig. \ref{fig2}. 
Two different behavior of the trajectories can be distinguished: motion toward higher or lower concentrations.}\label{fig3}
\end{figure}
\begin{figure}
  \centerline{\includegraphics[width=.90\columnwidth]{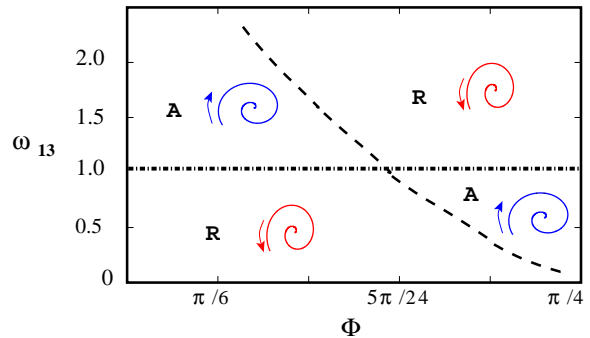}}% Images in 100% size
  \caption{(color online). 
Different behavior of the trajectories in a central concentration categorized in a phase diagram.
The horizontal axis $\Phi$ stands for a geometrical variable of the system and the vertical 
axis shows $\omega_{13}$, the anisitropy parameter that makes the individual swimmer an active system. 
Spiral trajectories going toward (backward) the region with higher concentration where is located in 
the center, are denoted by A (R).
Parameters have the same values as Fig. \ref{fig4}.
}
\label{fig4}
\end{figure}

Now let us consider a fluid medium occupied by a very low concentration of chemichal attractants given 
by $\rho({\bf x})$. The low concentration assumption makes sure that the hydrodynamic properties of the 
medium are not affected by this chemical. This chemical can derive the system to a nonequilibrium 
case by affecting the internal conformational 
changes. Inspired by the chemotactic navigation in biological microorganisms, we assume that there is 
a density sensing mechanism in the system. 
In the signaling network of the chemotactic systems, a dynamical mechanism is capable to produce an 
output signal that depends on the time history of the stimulating properties. Local concentration of the 
chemotoattractants is the stimulating properties in this case. 
Mathematics of such a generic adaptation mechanism can be modeled by a simple relaxational process \cite{frankchem,baraki}.  In this model an internal adaptation variable $u(t)$ 
couples to a signaling function $s(t)$ through the following equations:
\begin{equation}
\sigma\dot{s}=\rho u-s,~~~~~~~\mu\dot{u}=u(1-s),
\end{equation}
where the adaptation variable $u$, measures the dynamical sensitivity with the time scale of 
adaptation given by $\mu$. 
The relaxational process in a time scale controlled by $\sigma$ will produce a stimulus 
$s(t)$ that can affect the dynamical variables of the conformational changes which are already 
defined by $\omega_{ij}$. As an example we consider a case where the adaptation mechanism control 
only one of the transition rates. 
Here we choose $\omega_{21}=s(t)\omega_{12}$ and set all the other rates including $\omega_{12}$, 
to $\omega_0$. 
For a uniform profile of the concentration, the system reaches a steady state with $s=1$ that is not moving. If we set an 
asymmetry in one of the other internal conformational jumps, for example $\omega_{13}\ne\omega_{31}$, the steady state solution is a circling path. 
Now we can investigate the responce of a circling swimmer, embedded in a gradient of chemical concentration.
First we consider a linear gradient of chemical concentration given by $\rho({\bf x})=a_1+a_2x$. Fig. \ref{fig2} 
shows the trajectory of the swimmer for two different vales of the rates and the geometrical characteristics of 
the swimmer. The results can be summarized as follow: in general, nonuniform chemical concentration initiates a drift and the path looks like a drifting circle moving to the region with either higher 
or lower concentration value. For a constant asymmetry in one of the conformational changes, here $\omega_{13}-\omega_{31}>0$, there 
is a critical value for $\Phi_c$. Swimmer with geometrical structure that corresponds to 
$\Phi>\Phi_c$ ($\Phi<\Phi_c$) move to the region with higher (lower) concentration. 
%Higher $\Phi$ makes a better alignment of the direction of the drift 
%and the  direction of the gradient. 
Interestingly all the results are independent of the initial orientation of the swimmer and are not sensitive to the adaptation parameteres.

Motion in central distribution of the concentration has the same characteristics as linear gradient. Fig. \ref{fig3} shows the results for motion in a central gradient described 
by $\rho({\bf x})=a_3/|{\bf x}|$, with $a_3>0$. 
The trajectory in this case is a drifting circle along a spiral path. By changing the parameters we can achieve a swimmer with spiral trajectory that is moving either toward  or away from the center. 
A phase diagram for different kinds of the possible states is shown in Fig. \ref{fig4}. 
In this phase diagram attractive states into the center (region with high concentration) and repulsive states from  the center are denoted by labels $(A)$ and $(R)$ respectively. The horizontal axis $\Phi$, shows the geometrical variable of the system and the vertical 
axis $\omega_{13}$, is the asymmetric parameter that makes the individual swimmer an active system. 
This picture shows that for an active swimmer (a system that can move in a uniform concentration), 
by adjusting the parameters it is always possible to find states which can navigate to the correct direction.

In a concentration gradient, the force due to the nonequilibrium concentraton is also 
important \cite{osmotic}. Here we have neglected the effects due to this force. 
A dimensional analysis shows that for  satisfying this condition, the 
following criterion must hold: $|\nabla\rho|\ll\omega_0(a\epsilon\eta)/(R^3k_BT)$, where $\omega_0$ is 
a typical value for the rates of conformational changes.
In conclusion, we have introduced a micron scale system that uses the physics of 
chemotaxis in bacteria and navigate along a preferred direction into a source. The system is not 
sensitive to the adaptation variables. This two dimensional 
model can be readily generalized to three dimensional case. 

\acknowledgements
I acknowledge F. Mohammad-Rafiee for reading the manuscript, and financial support from the MPIPKS.

\end{document}